\documentclass[pdflatex,sn-mathphys-num]{sn-jnl}

\usepackage{graphicx}%
\usepackage{multirow}%
\usepackage{amsmath,amssymb,amsfonts}%
\usepackage{amsthm}%
\usepackage{mathrsfs}%
\usepackage[title]{appendix}%
\usepackage{xcolor}%
\usepackage{textcomp}%
\usepackage{manyfoot}%
\usepackage{booktabs}%
\usepackage{algorithm}%
\usepackage{algorithmicx}%
\usepackage{algpseudocode}%
\usepackage{listings}%

\theoremstyle{thmstyleone}%
\theoremstyle{thmstyletwo}%

\theoremstyle{thmstylethree}%

\begin{document}

\title[Article Title]{Encoding of Demographic and Anatomical Information in Chest X-Ray-based Severe Left Ventricular Hypertrophy Classifiers}


\author*[1]{\fnm{Basudha} \sur{Pal}}\email{bpal5@jhu.edu}

\author[1,2]{\fnm{Rama} \sur{Chellappa}}\email{rchella4@jhu.edu}

\author[3,4]{\fnm{Muhammad} \sur{Umair}}\email{mu2331@cumc.columbia.edu}

\affil*[1]{\orgdiv{Department of Electrical and Computer Engineering}, \orgname{Johns Hopkins University}, \orgaddress{ \city{Baltimore}, \state{MD}, \country{USA}}}

\affil[2]{\orgdiv{Department of Biomedical Engineering}, \orgname{Johns Hopkins University}, \orgaddress{ \city{Baltimore}, \state{MD}, \country{USA}}}

\affil[3]{\orgdiv{The Russell H. Morgan Department of Radiology and Radiological Sciences}, \orgname{The Johns Hopkins Hospital}, \orgaddress{ \city{Baltimore}, \state{MD}, \country{USA}}}

\affil[4]{\orgdiv{Department of Radiology}, \orgname{Columbia University Irving Medical Center}, \orgaddress{ \city{New York}, \state{NY}, \country{USA}}}

\abstract{While echocardiography and MRI are clinical standards for evaluating cardiac structure, their use is limited by cost and accessibility. We introduce a direct classification framework that predicts severe left ventricular hypertrophy from chest X-rays, without relying on anatomical measurements or demographic inputs. Our approach achieves high AUROC and AUPRC, and employs Mutual Information Neural Estimation to quantify feature expressivity. This reveals clinically meaningful attribute encoding and supports transparent model interpretation.}

\maketitle
                                                                

Early detection of left ventricular (LV) abnormalities such as severe hypertrophy (SLVH) is essential for cardiovascular risk management \cite{heidenreich20222022}. Conventionally, echocardiography remains the clinical standard for assessing left ventricular (LV) structure \cite{cheitlin1997acc}. However, its use is typically confined to individuals with a high pretest probability, in part due to reliance on specialized equipment and operator expertise \cite{heidenreich1999echocardiography}. While cardiac magnetic resonance imaging (MRI) provides accurate and reproducible assessments of myocardial structure and function \cite{khurshid2021deep}, its cost, limited availability, and scanning time restricts routine population-level screening.
Chest X-rays (CXRs) are widely available in clinical practice, noninvasive, inexpensive, and commonly used as a first-line imaging modality \cite{gurney1995chest}. Although chest X-rays (CXRs) lack three-dimensional detail and dynamic imaging capabilities for comprehensive cardiac anatomical assessment, recent deep learning models have demonstrated their ability to extract clinically meaningful information and estimate echocardiographic indices such as IVSDd, LVIDd, and LVPWDd \cite{bhave2024deep}.

However, directly classifying SLVH from CXRs remains challenging due to the subtle nature of phenotypic changes and 3-dimensional overlap of the modality. Consequently, prior work \cite{bhave2024deep} adopts an indirect approach of first performing regression to estimate intermediate anatomical variables such as IVSDd, LVIDd, and LVPWDd, followed by thresholding to determine SLVH status. Additionally, since, clinical thresholds for defining SLVH depend on age and gender, demographic information is explicitly provided to guide a learnable thresholding process and improve predictive alignment with clinical guidelines \cite{lang2015recommendations}. While this strategy can improve predictive performance, it also introduces architectural and conceptual drawbacks. First, it creates vulnerability to error cascading, where inaccuracies in anatomical regression propagate through the model and compromise final classification. Second, explicitly including demographic variables can result in confounding, as these attributes may serve as proxies for the outcome rather than truly independent predictors, thus reducing model transparency and interpretability. These issues make it harder to understand model behavior and raise concerns for clinical deployment, where clarity and robustness are essential. Beyond the risk of error propagation and confounding, from a representation learning perspective, it is preferable to train models directly on the end task \cite{bengio2013representation}. A dedicated classification model should learn features for discriminating between SLVH-positive and SLVH-negative cases, potentially offering better alignment with the final diagnostic objective and fewer sources of modeling error while implicitly encoding relevant clinical and demographic attributes. 

In this work, we propose a simplified yet effective alternative. We introduce a direct classification framework that predicts SLVH status (present or absent) from chest X-rays alone, without relying on intermediate structural predictions or demographic inputs. In this way, we avoid the risks associated with explicitly including anatomical or demographic variables, such as confounding and error propagation, while still aiming for a model that captures clinically meaningful patterns. To verify that our classifier remains aligned with relevant clinical attributes despite not receiving them as inputs, we apply Mutual Information Neural Estimation (MINE) \cite{belghazi2018mutual} to quantify the relationship between internal feature representations and relevant attributes such as age, sex, IVSDd, LVIDd, and LVPWDd. MINE estimates mutual information by training a neural network discriminator to distinguish joint from independent samples using the Donsker–Varadhan representation \cite{donsker1975asymptotic}, making it tractable in high-dimensional settings. This framework has been used to reveal how sensitive attributes are encoded in domains such as face recognition \cite{dhar2020attributes} and person reidentification \cite{pal2025quantitative}. We adapt it here to examine how clinical and demographic information is encoded across early, mid, and late layers in SLVH classifiers trained without access to those attributes. The resulting score, which we refer to as expressivity, reflects the degree to which each attribute is entangled in the learned representation. Further methodological details are provided in section \ref{secm}. 

This analysis quantifies the extent to which internal feature representations encode key variables such as age, sex, and anatomical measurements like IVSDd and LVPWDd. Although these factors are not part of the model input, they are strongly associated with SLVH and should be reflected in the features learned by an effective model. MINE helps us confirm that the model attends to this underlying structure implicitly, supporting both interpretability and clinical trust without compromising design efficiency or robustness. Our contributions are as follows:
\begin{itemize}
\item Modeling: We present a direct SLVH classification framework using both convolutional as well as transformer backbones with only chest X-ray images, removing reliance on anatomical regressors and demographic inputs. This improves generalizability and streamlines model design.
\item Evaluation: We address limitations in prior work by constructing a balanced subset of the CheXchoNet dataset, enabling improved discriminative performance and reliable benchmarking. Performance is assessed using AUROC and AUPRC.
\item Interpretability: We apply MINE to estimate mutual information between internal features and clinical attributes, enabling quantitative analysis of attribute encoding without requiring explicit supervision. This supports a more interpretable and clinically aligned deployment of deep learning models.
\end{itemize}
To our knowledge, this is the first study to propose a direct classification framework for detecting SLVH from chest X-rays, bypassing the need for intermediate anatomical modeling or demographic inputs. Moreover, we are the first to introduce MINE as a tool for analyzing internal representations in deep learning based cardiac imaging. Beyond the specific application to SLVH classification, MINE represents a generalizable framework for quantifying feature–attribute relationships in deep learning models. Its ability to uncover clinically meaningful correlations makes it particularly well suited for advancing interpretability in cardiac imaging and other domains of clinical computer vision.

\begin{figure}[!htbp]
\centering
\includegraphics[width = \textwidth]{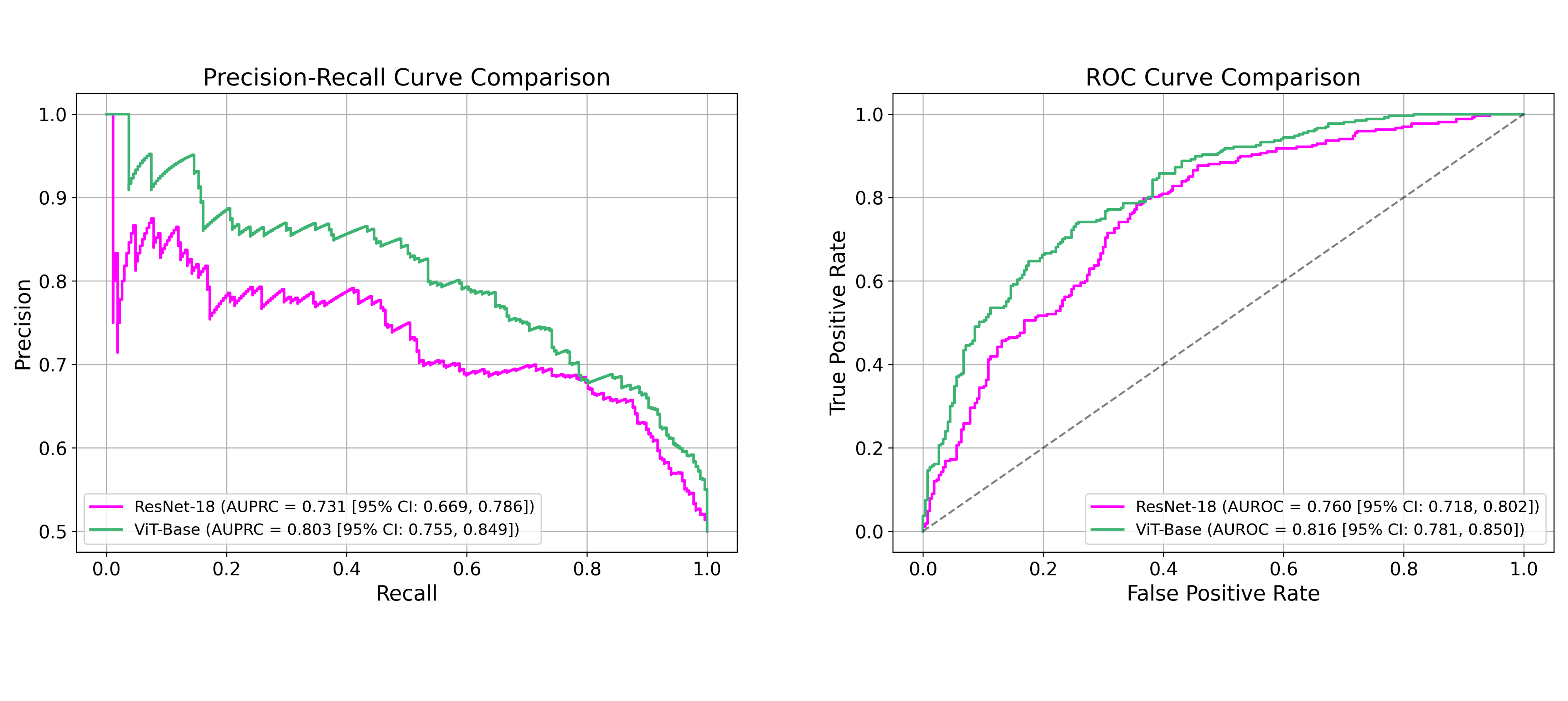}
\caption{SLVH classification performance shown via \textbf{Left:} PR and \textbf{Right:} ROC curves for ResNet and ViT backbones.}
\label{fig2}
\end{figure}

\begin{figure}[!htbp]
\centering
\includegraphics[width = \textwidth]{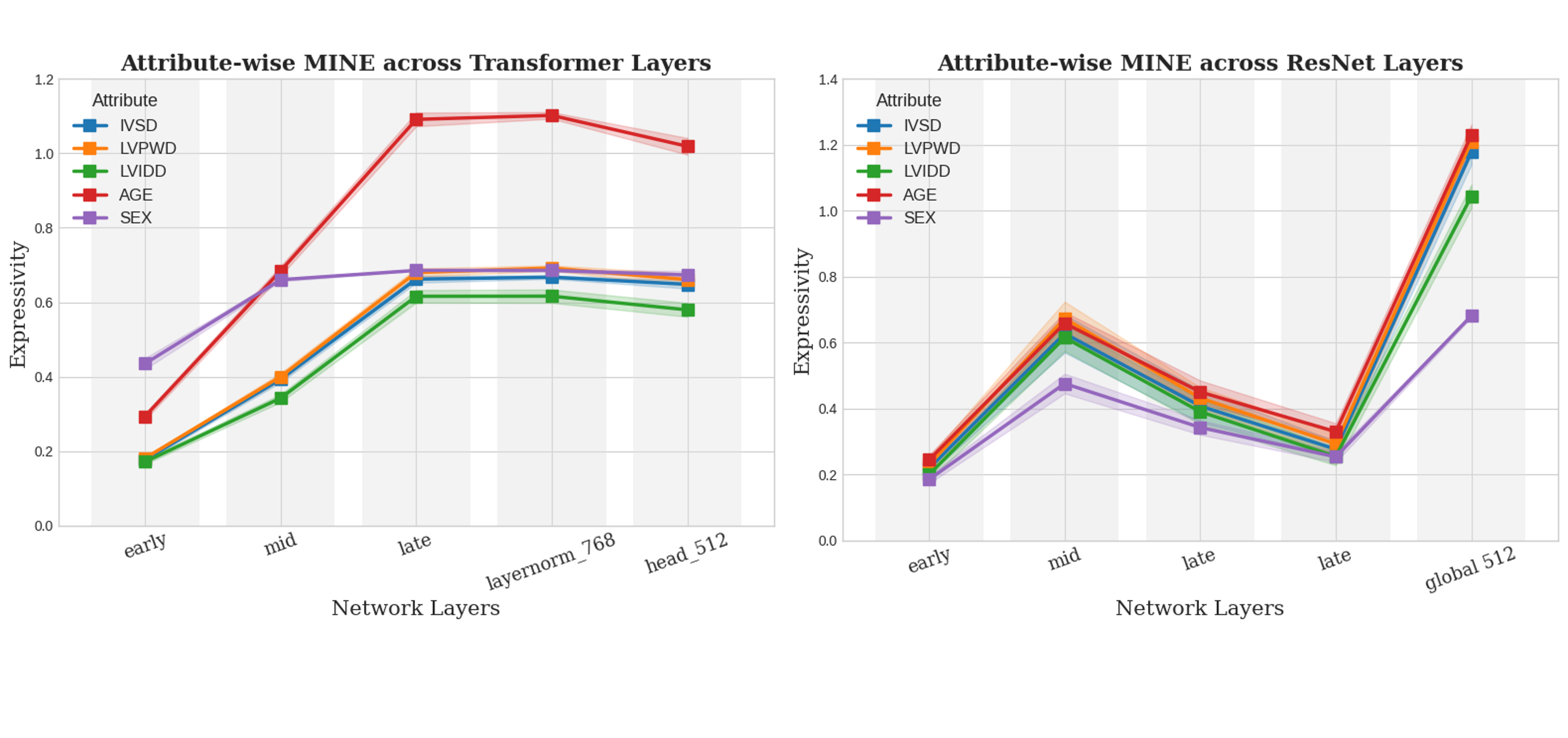}
\caption{Attribute-wise expressivity across layers for \textbf{Left:} ViT and \textbf{Right:} ResNet backbones. The standard deviation of each of the curves is denoted by the shaded region.}
\label{mine}
\end{figure}
We demonstrate that SLVH can be directly classified from chest X-rays using deep neural network backbones without reliance on intermediate anatomical regressors or demographic inputs. The current baseline \cite{bhave2024deep}, which predicts SLVH by regressing echocardiographic measurements followed by thresholding, achieves high AUROC (0.79 [95\% CI: 0.76–0.81]) but suffers from poor AUPRC (0.19 [95\% CI: 0.15–0.22]) due to class imbalance. Fine-tuning pretrained models on our curated class-balanced dataset yields strong performance across both convolutional and transformer-based architectures. Specifically, our pretrained Vision Transformer (ViT) backbone achieves the best AUROC of 0.816 [95\% CI: 0.781, 0.850], and an AUPRC of 0.803 [95\% CI: 0.755, 0.849], while a ResNet-18 backbone achieves AUROC of 0.760 [95\% CI: 0.718, 0.802], and an AUPRC of 0.731[95\% CI: 0.669, 0.786] as seen in Fig ~\ref{fig2}. These results suggest that the ViT model better captures features relevant to SLVH, likely due to its ability to model long range dependencies through global self-attention.
\begin{table}[ht]
\centering
\begin{tabular}{lcc}
\toprule
\textbf{Model} & \textbf{AUROC [95\% CI]} & \textbf{AUPRC [95\% CI]} \\
\midrule
Baseline (Bhave et al. \cite{bhave2024deep}) & 0.79 [0.76–0.81] & 0.19 [0.15–0.22] \\
ResNet-18 backbone & 0.76 [0.72, 0.80] & 0.73 [0.67, 0.79] \\
ViT backbone & \textbf{0.82} [0.78, 0.85] & \textbf{0.80} [0.76, 0.85] \\
\bottomrule
\end{tabular}
\caption{Comparison of SLVH classification performance across models. While the baseline model achieves a relatively high AUROC, its AUPRC is limited due to class imbalance. Fine-tuned ResNet-18 improves AUPRC, and the Vision Transformer achieves the best performance on both AUROC and AUPRC (in bold).}
\label{tab:slvh-performance}
\end{table}

To examine whether the models internally encode clinically meaningful signals, we assess what information they retain about relevant anatomical and demographic attributes. Using MINE, we evaluate the expressivity of five variables: IVSDd, LVPWDd, LVIDd, age, and sex across different network layers as seen in Fig \ref{mine}. In ResNet-18, expressivity remains relatively low throughout intermediate layers but increases sharply at the final global representation. In contrast, the ViT model shows a more gradual increase in expressivity across successive attention blocks, consistent with its architecture that enables progressive integration of contextual features.

Across both models, age and sex show high expressivity in deeper layers. While this may raise questions regarding fairness, it also reflects the clinical relevance of these attributes, since SLVH diagnosis often depends on age specific and sex specific criteria. Among anatomical markers, IVSDd and LVPWDd are identified as key indicators of myocardial thickening and exhibit rising expressivity with depth, confirming that both models focus on structurally meaningful features. In contrast, LVIDd consistently shows low expressivity throughout the network. This observation aligns with clinical understanding, as LVIDd is more strongly associated with ventricular dilation than with hypertrophy. Taken together, these findings reveal a consistent ordering of attribute expressivity: age \(>\) sex \(>\) IVSDd \(>\) LVPWDd \(>\) LVIDDd, observed across both model architectures. This pattern aligns with clinical reasoning, where demographic variables inform diagnostic thresholds and septal or posterior wall thickness serves as a primary indicator of hypertrophy. The ability of the models to reflect this prioritization despite being trained only on binary SLVH labels illustrates the effectiveness of the models. Moreover, it demonstrates the utility of MINE-based analysis for uncovering clinically meaningful structure in the learned representations.

 In future work, LVIDd could be used as a control variable to evaluate suppression strategies, while age and sex may require partial suppression or conditional modeling to balance fairness with clinical utility. These results support the effectiveness of our direct classification framework, which avoids the cascading errors and complexity of multi-stage pipelines and show that MINE based expressivity analysis enables detailed interpretation of clinically relevant representations without requiring explicit supervision.

\begin{figure}[!htbp]
\centering
\includegraphics[width = \textwidth]{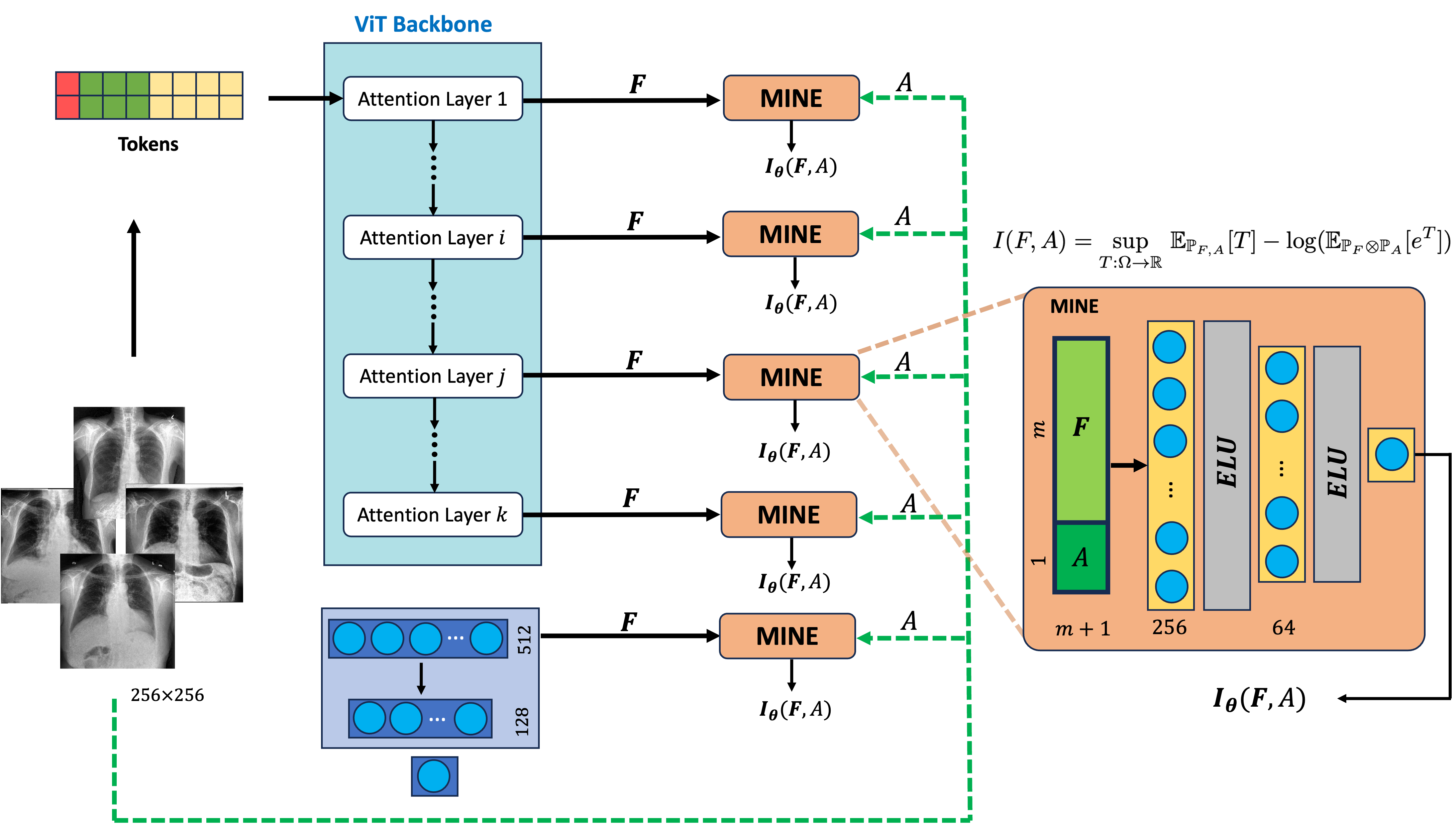}
\caption{MINE-based expressivity pipeline for a ViT backbone used in SLVH classification. Feature representations \( F \in \mathbb{R}^m \) are extracted from a designated transformer layer and concatenated with scalar attribute values \( A \) (such as age, sex, IVSd, LVPWd, or LVIDd), resulting in \( (m + 1) \)-dimensional inputs. These inputs are passed through a two-layer multilayer perceptron (MLP) within the MINE module, which is trained to estimate mutual information and quantify the degree to which each attribute is encoded in the model’s internal features.}
\label{fig1}
\end{figure}

\section{Methods}\label{secm}

\subsection{Dataset and Preprocessing}

We use the CheXchoNet dataset \cite{bhave2024deep}, which pairs chest X-rays (CXRs) with structural measurements derived from echocardiography. Previous models \cite{bhave2024deep} performed regression followed by thresholding to predict severe left ventricular hypertrophy (SLVH), but demonstrated limited discriminative performance due to pronounced class imbalance. To address this, we construct class-balanced subsets by sampling equal numbers of SLVH-positive and SLVH-negative cases while preserving the original train, validation, and test split proportions. The final dataset includes 11,190 CXRs (5,595 per class) from 6,021 patients for training, 658 CXRs (329 per class) from 361 patients for validation, and 534 CXRs from 310 patients for testing. All images are resized to 256 $\times$ 256 and normalized using ImageNet preprocessing statistics.

\subsection{Classification Architectures} 
We evaluate two deep learning backbones for direct SLVH classification from CXRs:
\begin{itemize}
\item ResNet-18: A pretrained ResNet-18 \cite{he2016deep} is fine-tuned end-to-end. The backbone is appended with fully connected layers (FC-128 $\rightarrow$ FC-64 $\rightarrow$ FC-1), each followed by ReLU activation and dropout (p = 0.3). The model is trained using the Adam optimizer \cite{kingma2014adam} with a learning rate = 1e-4 and weight decay = 1e-5.

\item Vision Transformer (ViT): A ViT encoder pretrained on chest X-rays using a masked autoencoding strategy \cite{xiao2023delving} is initialized with pretrained weights and paired with a trainable classification head (MLP: 512 $\rightarrow$ 128 $\rightarrow$ 1). The encoder remains frozen while the head is trained using linear warmup followed by cosine learning rate decay. Figure \ref{fig1} is a comprehensive block diagram for integrating MINE with a ViT backbone.

\end{itemize}
Both models are trained using binary cross-entropy loss, and the checkpoint with the highest validation AUROC is selected for evaluation.

\subsection{Expressivity Analysis using MINE}

To analyze what clinical information is captured by the models despite being trained without access to demographic or anatomical variables, we evaluate the extent to which learned feature representations encode clinically meaningful attributes. We refer to this capacity as expressivity, measured as the mutual information (MI) between internal features and a given attribute.

Mutual information quantifies the statistical dependence between variables. In our case, it captures how much knowing the model's internal representation reduces uncertainty about an attribute such as age or myocardial wall thickness. Since estimating MI in high-dimensional spaces is challenging, we use MINE \cite{belghazi2018mutual}, which approximates MI using a neural network trained to distinguish real (paired) feature-attribute examples from randomly shuffled (unpaired) ones. MINE optimizes the Donsker-Varadhan lower bound as seen in Equation \ref{eqn1}:

\begin{equation}
    I(F; A) \geq \mathbb{E}_{P_{FA}}[T_\theta(f, a)] - \log \mathbb{E}_{P_F \otimes P_A}[e^{T_\theta(f, a)}]
    \label{eqn1}
\end{equation}

Here, $T_\theta$ is a discriminator network with parameters $\theta$, trained to output higher scores for paired inputs drawn from the joint distribution $P_{FA}$, and lower scores for mismatched pairs drawn from the product of marginals $P_F \otimes P_A$.

\subsection{Implementation and Procedure} 

We implement $T_\theta$ as a multilayer perceptron with two hidden layers of 256 and 64 units, using ELU activation. The network is trained using the Adam optimizer (learning rate = 1e-3, batch size = 100), with weights initialized using Xavier initialization. To ensure robustness, we repeat the estimation over $M = 10$ random seeds and report the average expressivity.

For each attribute, we compute expressivity from multiple network layers (early, mid, and final), separately for ResNet-18 and ViT. Each image $x_i$ is passed through the network to extract its feature vector $f_i$. These are stacked to form a feature matrix $\mathbf{F} = [f_1, f_2, ..., f_n]^T$, which is concatenated with the corresponding attribute vector $\mathbf{A} \in \mathbb{R}^{n \times 1}$ to yield the input $\mathbf{X} = [\mathbf{F} | \mathbf{A}]$.

The MINE network is then trained to distinguish paired and unpaired inputs, and outputs a score indicating the level of dependency between the features and the attribute. This procedure is repeated over 10 random seeds to account for stochasticity in training, and the average MI is reported as the final expressivity score. In simple terms, we are asking how well the model's internal representations, trained only to predict SLVH, also reflect other relevant clinical signals even though it was never explicitly told about them. The steps of this procedure are outlined in Algorithm \ref{alg:expressivity}.

\begin{algorithm}[H]
\caption{Expressivity Computation on learnt representations}
\label{alg:expressivity}
\begin{algorithmic}[1]
\Require Layer \( L \), set of \( n \) images \( I \), attribute vector \( \mathbf{A} \in \mathbb{R}^{n \times 1} \)
\Ensure Expressivity measure
\State Initialize \( E \gets [ ] \) \Comment{To store expressivity values}
\State Extract features \( \mathbf{F} \gets [f_1, f_2, \dots, f_n]^T \) from \( L \) after a particular epoch for all \( i \in I \)
\State Concatenate the features and attributes: \( \mathbf{X} \gets [\mathbf{F} | \mathbf{A}] \) \Comment{Augmentation step}
\For{\( \text{iteration} = 1 \) to \( M \)}
    \State Initialize MINE network \( T_\theta \) based on the dimensions of \( \mathbf{X} \)
    \State Compute expressivity score: \( e \gets \text{MINE}(\mathbf{X}) \)
    \State Append score: \( E \gets E \cup \{e\} \)
\EndFor
\State \Return \( \text{Expressivity} \gets \text{Average}(E) \)
\end{algorithmic}
\end{algorithm}

We analyze the expressivity of five attributes: age, sex, IVSd, LVPWd, and LVIDd. These attributes are clinically relevant to SLVH diagnosis but are not used during training. Age and sex are drawn from demographic metadata, while structural measurements are derived from echocardiograms. Our findings reveal a consistent encoding hierarchy across both architectures: age $>$ sex $>$ IVSDd $>$ LVPWDd $>$ LVIDDd. This ordering aligns with clinical understanding, where age and myocardial wall thickness are key discriminators of hypertrophy.

\bibliography{sn-bibliography}

\end{document}